\documentclass[preprint,amsmath,amssymb,superscriptaddress,amsfonts,nofootinbib,showpacs]{revtex4-1}
\usepackage{graphicx}  
\usepackage{float}
\begin{document}

\preprint{preprint submitted to Phys. Rev. C}

\title{\textbf{Correlations between emission timescale of fragments and isospin dynamics 
in $^{124}$Sn+$^{64}$Ni and $^{112}$Sn+$^{58}$Ni reactions at 35 AMeV}}

\author{E.~De~Filippo}\email{defilippo@ct.infn.it}
\affiliation{INFN, Sezione di Catania, Italy}
\author{A.~Pagano}\affiliation{INFN, Sezione di Catania, Italy}
\author{P.~Russotto}
\affiliation{INFN, Laboratori Nazionali del Sud, Catania, Italy}
\affiliation{Dipartimento di Fisica e Astronomia, Univ. di Catania, Catania, Italy}
\author{F.~Amorini}
\affiliation{INFN, Laboratori Nazionali del Sud, Catania, Italy}
\affiliation{Dipartimento di Fisica e Astronomia, Univ. di Catania, Catania, Italy}
\author{A.~Anzalone}\affiliation{INFN, Laboratori Nazionali del Sud, Catania, Italy}
\author{L.~Auditore}\affiliation{INFN, Gr. Coll. di Messina and Dip. di Fisica, Univ. di Messina, Italy}
\author{V.~Baran}\affiliation{Physics Faculty, University of Bucharest, Romania}
\author{I.~Berceanu}\affiliation{National Institute of Physics and Nuclear Engineering "Horia Hulubei", Bucharest, Romania}
\author{B.~Borderie}\affiliation{Institut de Physique Nucl$\acute{e}$aire, CNRS/IN2P3, Universit$\acute{e}$ Paris-Sud 11, Orsay, France}
\author{R.~Bougault}\affiliation{LPC Caen, ENSICAEN, University of Caen, CNRS/IN2P3, Caen France}
\author{M.~Bruno}\affiliation{INFN, Sezione di Bologna and Dipartimento di Fisica, Univ. di Bologna, Italy}
\author{T. Cap} \affiliation{Faculty of Physics, University of Warsaw, Warsaw, Poland}
\author{G.~Cardella}\affiliation{INFN, Sezione di Catania, Italy}
\author{S.~Cavallaro}
\affiliation{INFN, Laboratori Nazionali del Sud, Catania, Italy}
\affiliation{Dipartimento di Fisica e Astronomia, Univ. di Catania, Catania, Italy}
\author{M.B.~Chatterjee}\affiliation{Saha Institute of Nuclear Physics, Kolkata, India}
\author{A.~Chbihi}\affiliation{GANIL (DSM-CEA/CNRS/IN2P3), Caen, France}
\author{M.~Colonna}\affiliation{INFN, Laboratori Nazionali del Sud, Catania, Italy}
\author{M.~D'Agostino}\affiliation{INFN, Sezione di Bologna and Dipartimento di Fisica, Univ. di Bologna, Italy}
\author{R.~Dayras}\affiliation{DAPNIA/SPhN, CEA-Saclay, France}
\author{M.~Di~Toro}
\affiliation{INFN, Laboratori Nazionali del Sud, Catania, Italy}
\affiliation{Dipartimento di Fisica e Astronomia, Univ. di Catania, Catania, Italy}
\author{J.~Frankland}\affiliation{GANIL (DSM-CEA/CNRS/IN2P3), Caen, France}
\author{E.~Galichet}\affiliation{Institut de Physique Nucl$\acute{e}$aire, CNRS/IN2P3, Universit$\acute{e}$ Paris-Sud 11, Orsay, France}
\author{W.~Gawlikowicz}
\affiliation{Cardinal Stefan Wyszy\'nski University, Warsaw, Poland}
\author{E.~Geraci}
\affiliation{Dipartimento di Fisica e Astronomia, Univ. di Catania, Catania, Italy}
\affiliation{INFN, Sezione di Catania, Italy}
\author{A.~Grzeszczuk}\affiliation{Institute of Physics, University of Silesia, Katowice, Poland}
\author{P.~Guazzoni}\affiliation{INFN, Sezione di Milano and Dipartimento di Fisica, Univ. di Milano, Italy}
\author{S.~Kowalski}\affiliation{Institute of Physics, University of Silesia, Katowice, Poland}
\author{E.~La~Guidara}\affiliation{INFN, Sezione di Catania, Italy}
\author{G.~Lanzalone}
\affiliation{INFN, Laboratori Nazionali del Sud, Catania, Italy}
\affiliation{``Kore'' Universit$\acute{a}$, Enna, Italy}
\author{G.~Lanzan\`o}\altaffiliation{\textbf{deceased}}\affiliation{INFN, Sezione di Catania, Italy}
\author{N.~Le~Neindre}\affiliation{LPC Caen, ENSICAEN, University of Caen, CNRS/IN2P3, Caen France}
\author{I. Lombardo}
\affiliation{INFN, Sezione di Napoli and Dipartimento di Fisica, Univ. di Napoli, Italy}
\author{C.~Maiolino}\affiliation{INFN, Laboratori Nazionali del Sud, Catania, Italy}
\author{M.~Papa}\affiliation{INFN, Sezione di Catania, Italy}
\author{E.~Piasecki}\affiliation{Heavy Ion Laboratory, University of Warsaw, Warsaw, Poland}
\affiliation{National Centre for Nuclear Research, Otwock-\'Swierk, Poland}
\author{S.~Pirrone}\affiliation{INFN, Sezione di Catania, Italy}
\author{R.~P{\l}aneta}\affiliation{M.Smoluchowski Institute of Physics, Jagellonian Univ., Cracow, Poland}
\author{G.~Politi}
\affiliation{Dipartimento di Fisica e Astronomia, Univ. di Catania, Catania, Italy}
\affiliation{INFN, Sezione di Catania, Italy}
\author{A.~Pop}\affiliation{National Institute of Physics and Nuclear Engineering "Horia Hulubei", Bucharest, Romania}
\author{F.~Porto}
\affiliation{INFN, Laboratori Nazionali del Sud, Catania, Italy}
\affiliation{Dipartimento di Fisica e Astronomia, Univ. di Catania, Catania, Italy}
\author{M.F.~Rivet}\affiliation{Institut de Physique Nucl$\acute{e}$aire, CNRS/IN2P3, Universit$\acute{e}$ Paris-Sud 11, Orsay, France}
\author{F.~Rizzo}
\affiliation{INFN, Laboratori Nazionali del Sud, Catania, Italy}
\affiliation{Dipartimento di Fisica e Astronomia, Univ. di Catania, Catania, Italy}
\author{E. Rosato}\affiliation{INFN, Sezione di Napoli and Dipartimento di Fisica, Univ. di Napoli, Italy}
\author{K.~Schmidt}\affiliation{Institute of Physics, University of Silesia, Katowice, Poland}
\author{K.~Siwek-Wilczy\'nska}\affiliation{Faculty of Physics, University of Warsaw, Warsaw, Poland}
\author{I.~Skwira-Chalot}\affiliation{Faculty of Physics, University of Warsaw, Warsaw, Poland}
\author{A.~Trifir\`o}\affiliation{INFN, Gr. Coll. di Messina and Dip. di Fisica, Univ. di Messina, Italy}
\author{M.~Trimarchi}\affiliation{INFN, Gr. Coll. di Messina and Dip. di Fisica, Univ. di Messina, Italy}
\author{G.~Verde}\affiliation{INFN, Sezione di Catania, Italy}
\author{M.~Vigilante}\affiliation{INFN, Sezione di Napoli and Dipartimento di Fisica, Univ. di Napoli, Italy}
\author{J.P.~Wieleczko}\affiliation{GANIL (DSM-CEA/CNRS/IN2P3), Caen, France}
\author{J.~Wilczy\'nski}\affiliation{National Centre for Nuclear Research, Otwock-\'Swierk, Poland}
\author{L.~Zetta}\affiliation{INFN, Sezione di Milano and Dipartimento di Fisica, Univ. di Milano, Italy}
\author{W. Zipper}\affiliation{Institute of Physics, University of Silesia, Katowice, Poland}

\date{\today}

\begin{abstract}
We present a new experimental method to correlate the isotopic composition of intermediate mass fragments (IMF) emitted at mid-rapidity in semi-peripheral collisions with the emission timescale: IMFs emitted in the early stage of the reaction show larger values of $<$N/Z$>$ isospin asymmetry, stronger angular anisotropies and reduced odd-even staggering effects in neutron to proton ratio $<$N/Z$>$ distributions than those produced in sequential statistical emission. All these effects support the concept of isospin ``migration'', that is sensitive to the density gradient between participant and quasi-spectator nuclear matter, in the so called neck fragmentation mechanism. By comparing the data to a Stochastic Mean Field (SMF) simulation we show that this method gives valuable constraints on the symmetry energy term of nuclear equation of state at subsaturation densities. An indication emerges for a linear density dependence of the symmetry energy. 
\end{abstract}

\pacs{25.70Mn, 25.70Pq}
\maketitle

\section{Introduction}
In the Fermi energy domain (15-100 A.MeV) semi-peripheral heavy-ion collisions are characterized by the presence of ternary or quaternary reactions leading to the formation of a binary system with excited projectile-like (PLF) and target-like (TLF) fragments strongly correlated with one or more IMFs and light particles in the exit channel. The emission pattern of these reactions has shown that light charged particles (LCP, Z$\le$2) and IMFs (Z$\ge$3) are not entirely described by the statistical decay of the PLF and TLF. In particular, light IMFs (Z$\le$10) are emitted preferentially towards mid-rapidity region (intermediate between PLF and TLF rapidities). Their velocity distributions display typical forward-backward asymmetry in the invariant cross section $d^2\sigma/v_\perp dv_\perp dv_\parallel$ (see, for example, Fig. 1e) in section II), indicating their dynamical origin \cite{hud05,dit06,pia06}.
The term ``neck fragmentation'' is commonly used for such a type of events,
because a neck-like structure is predicted to be formed between the two main residues (PLF and  TLF), in the early stages of the reaction.
Transport model simulations \cite{bar04} have shown that IMFs in the neck region are formed in dilute matter in contact with the PLF and TLF residues. However, the probability to produce fragments at mid-rapidity depends on different variables, such as impact parameter, beam energy and isospin asymmetry \cite{luk03}. The effective interactions driving transport phenomena of neutrons and protons through the neck have been related to the slope (density gradient) and magnitude (isospin gradient) of the symmetry potential term of the Equation of State (EOS). Thus, heavy ion collisions with projectiles and targets with different isospin asymmetries have been studied to probe the density dependence of the symmetry term of the EOS (``asy-eos'') that is an important ingredient for nuclear structure and astrophysical phenomena \cite{hor01,bot10}. Different observables, mainly based upon measurement of the neutron to proton (N/Z) ratio of reactions products have shown sensitivity to the 
density dependence of the simmetry energy like isospin diffusion and equilibration \cite{tsa09,sun10,gal09}, neutron to proton ratio measurements \cite{fam06}, transverse collective flow \cite{koh11}. 

Neutron enrichment of LCPs and IMFs in the mid-rapidity region has been observed experimentally (see for example ref. \cite{koh11} and references therein).
An excess of neutron (isospin migration) towards the dilute low-density region is predicted by transport simulations. This is determined by the slope of the density dependence of the symmetry energy,
providing a drift contribution to the isospin transfer, thus producing neutron rich IMFs at mid-rapidity \cite{riz08,dit10,bar05}. Such neutron enrichment of the neck-like IMFs has never consistently been used to pin down the time scale of the reaction dynamics and to constrain the theoretical simulations vs. the isospin asymmetry. Indeed, a neutron enrichment of the neck region could be also affected by phenomena not strictly related to the isospin dynamics, like proximity effects in the decay of PLF and TLF \cite{pia06}, or effects of reduced size of neck structures and the persistence of neutron-rich matter at the surface \cite{dem96,sob00}. As shown in \cite{lom10} the isospin drift and isospin diffusion can simultaneously compete to characterize the midrapidity and projectile residue emission.

In ref \cite{wil05,def05,def09,pag04} a new method based on a three body analysis of fragment-fragment relative velocities has been introduced to calibrate the timescale of IMFs emission in semi-peripheral collisions, thus probing the dynamics and chronology of fragment formation.

In this paper, the time scale (chronology) of IMFs originating in ternary reactions is correlated with their isospin content by inspecting typical dynamical observable, such as the degree of alignment and relative velocity correlations. The results are compared in the last section to predictions of microscopic transport calculations for the neutron rich system. 

\begin{figure}
\includegraphics[width=0.60\textwidth]{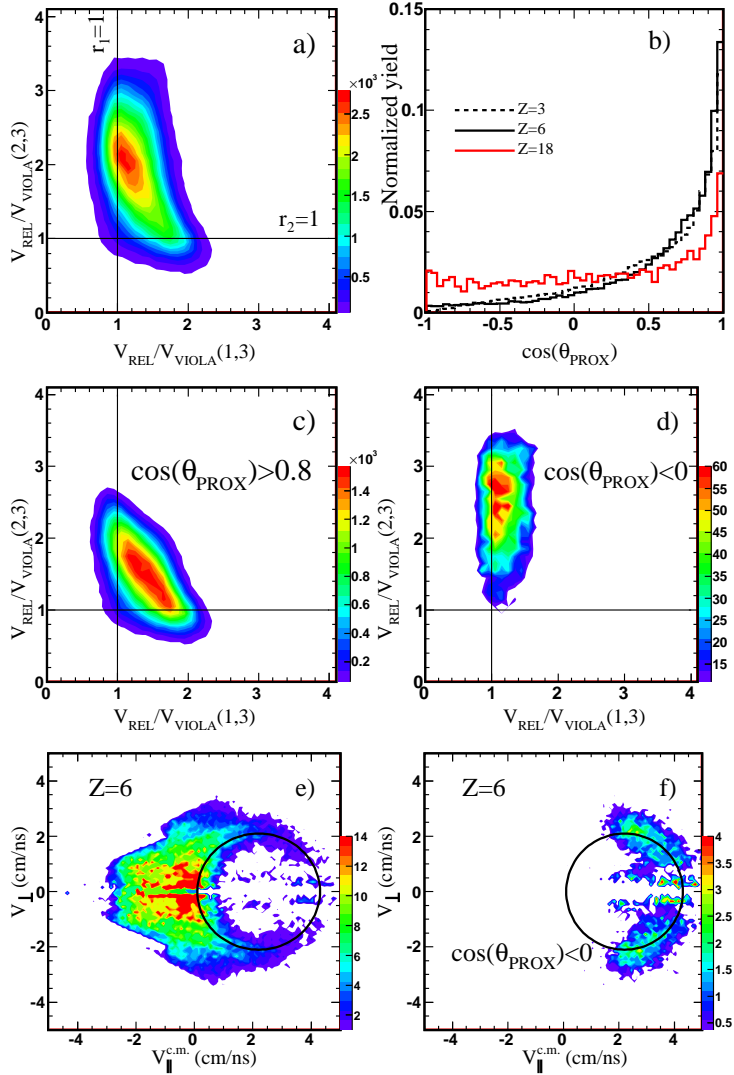}
\caption{(Color online) For the $^{124}$Sn+$^{64}$Ni reaction and M$\le6$ a) correlations between relative velocities V$_{REL}$/V$_{Viola}$ of the three biggest fragments of the event for IMFs (3$\le$Z$\le$18; b) distribution of $\cos(\theta_{PROX})$ for Z=3 (dotted line) Z=6 (thick line) and Z=18 (thin line);  c) as a) with the condition $\cos(\theta_{PROX})>$0.8; d) as a) with the condition $\cos(\theta_{PROX})<$0; e) Invariant cross-section for Z=6 IMFs. f) as e) with the condition $\cos(\theta_{PROX})<$0; the circumference shows a Coulomb ridge centered at PLF source velocity and 2.1 cm/ns radius.}
\label{fig1}
\end{figure}

\section{Experimental details and results}
The experiment was performed at the INFN-LNS Super-Conductive Cyclotron of Catania (Italy), bombarding thin ($\approx$ 300$\mu$g/cm$^2$) self-supporting $^{64}$Ni and $^{58}$Ni targets with 
35 A.MeV $^{124}$Sn and $^{112}$Sn beams, respectively.
Reaction products were detected with the forward part of the 4$\pi$ multi-detector CHIMERA \cite{pag04} that is constituted by 688 Si($\approx$300 $\mu m$)-CsI(Tl) telescopes over a total of 1192, arranged in 18 rings and covering the angular range between 1$^o$ and 30$^o$. 
Isotopic identification was obtained up to Oxygen by $\Delta E-E$ technique. Details about the 
experimental methods are discussed in \cite{pag04,len02}. In our reverse kinematics condition the angular coverage represents almost 85\% of the c.m. solid angle for the reaction channels under study.
Complete events (at least 70\% of the total charge and total parallel momentum of the colliding systems) were analyzed. Semi-peripheral collisions were selected gating on the total charged products multiplicity $M$. Here only events with M$\le6$ will be considered. These events correspond to a reduced impact parameter b$_{red}$ $\approx$ b/b$_{max}$ $\ge$ 0.7, determined using the Cavata approximation 
\cite{cav90} as explained in ref. \cite{rus10}. As shown in Fig. 2 of ref \cite{def05}, favorable conditions of reverse kinematics and capability of the CHIMERA device to detect fragments in a broad range of kinetic energies (including the slow moving target-like residues) greatly facilitate the distinction of PLF,TLF and IMFs in ternary events. 

In order to evaluate the timescale of fragment formation we extended the method quantitatively 
described in ref. \cite{wil05,def05}. The three biggest fragments in each event were sorted
according to the decreasing value of their parallel velocity V$_\parallel$ along the beam direction
(1=fastest fragment, 2=slowest fragment, 3=intermediate velocity fragment) and the fragment-fragment relative velocities, V$_{REL}$(1,3) and V$_{REL}$(2,3), were calculated. The fragment labels 1,2 and 3 correspond to the PLF, TLF residues and the IMF fragment, respectively, mentioned on ref. \cite{def05}. In fact, ranking the fragments according to their parallel velocity gives the simplest way to extend the present study towards the most dissipative collisions, associated with higher values of the IMFs multiplicity \cite{rus09}. The relative velocities are normalized to the one corresponding to the Coulomb repulsion, as given by the Viola systematics, V$_{Viola}$(i,3) (i=1,2) \cite{hin87}.
The correlation for fragments with charge 3$\le$Z$\le$18 between the two relative velocities r$_1$=V$_{REL}$/V$_{Viola}$(1,3) and r$_2$=V$_{REL}$/V$_{Viola}$(2,3) is shown in Fig. 1a): the values r$_1$=1 and r$_2$=1 correspond to sequential decay of IMFs from a PLF and TLF respectively; values of r$_1$ and r$_2$ simultaneously larger than unity indicate IMFs of dynamical origin (prompt ternary division). 
Timescale calibration of Fig. 1a) was done in ref. \cite{wil05,def05} using a three-body collinear Coulomb trajectory calculation,
showing a well defined chronology: light IMFs are emitted either on a short timescale (within 50 fm/c) or sequentially ($>$120 fm/c) after the re-separation of the binary PLF-TLF system. This result has been reproduced
by different transport reaction simulations like Stochastic Mean Field SMF \cite{bar04} and CoMD-II Constrained Molecular Dynamics model \cite{pap07} and it is in agreement with simulations based on similar three-body Coulomb trajectory calculations \cite{pian02}. It is interesting to note that antisymmetryzed molecular calculations (AMD) \cite{hud06} for semi-peripheral reactions also predicts the formation of cluster at mid-velocity on a fast timescale, within 90 fm/c, due to the inset of shape and density fluctuations.

In order to characterize the dynamical emission v.s. isospin degree of freedom, the quantity
$\cos(\theta_{PROX})$ was evaluated. $\theta_{PROX}$ gives the angle between the separation axis $\mathbf n_S=(\mathbf V_{PLF-IMF}^{c.m.} - \mathbf V_{TLF})$ (relative velocity between TLF and PLF-IMF center of mass) and the break-up axis $\mathbf n_F=(\mathbf V_{PLF}-\mathbf V_{IMF})$ (relative velocity between PLF and IMF oriented from the light to the heavy fragment), i.e.,
$\cos(\theta_{PROX}) =
\frac{\mathbf n_S \cdot \mathbf n_F}{\vert \mathbf n_S\vert\vert \mathbf n_F\vert}$.
\begin{figure}
\includegraphics[width=0.60\textwidth]{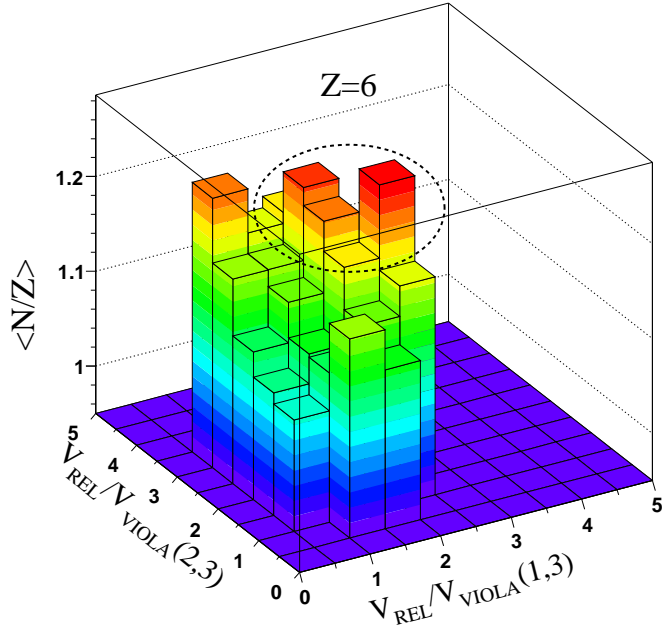}
\caption{(Color online) For the $^{124}$Sn + $^{64}$Ni reaction and M$\le6$: $<$N/Z$>$
for charge Z=6 for different bins in the r$_1$-r$_2$ plane. The dashed contour, projected in the r$_1$-r$_2$ plane gives a schematical selection of the zone with values of r$_1$ and r$_2$ 
simultaneously larger than unity.}
\label{fig2}
\end{figure}
Notice that this definition is slightly different from the one used in \cite{Boc20,mci10} 
since it requires the crucial detection of a TLF fragment, as it is in our case. 
Recently, contemporary TLF-PLF detection at Fermi energies has produced
important advances in the fields of mass-energy transport
phenomena \cite{man04} and decay of very heavy nuclei \cite{skw08}.
Fig. 1b) shows the $\cos(\theta_{PROX})$ distribution for Z=3, Z=6 and Z=18 IMFs charges respectively. The strong enhancement of the distribution for $\cos(\theta_{PROX})>$0.8 as seen in Fig. 1b) indicates
a clear contribution of dynamical emission. Notice the strong tendency
(for $cos(\theta_{PROX})\approx$1) to a backward IMF emission in a strict aligned configuration along the TLF-PLF separation axis. The enhancement for the Z=18 charge is mainly due to the onset of 
dynamical fission of the projectile as shown in \cite{rus10}.
By setting the condition $cos(\theta_{PROX})<$0 (forward emission) in Fig.1a) we obtain the pattern of Fig. 1d), where the events populate the region around the axis where r$_1$=1, as expected for a sequential decay of the IMF from the PLF source. In contrast, in Fig 1c) the correlation of Fig. 1a) is obtained by selecting events with the conditions $\cos(\theta_{PROX})>$0.8 (backward emission), showing
events concentrated near the diagonal, as expected for the dynamical emission of fragments.  
These results are further illustrated by comparing  the invariant cross-section distributions of Fig. 1e) and Fig. 1f) as a function of V$_\parallel$-V$_\perp$ velocities with respect to the beam axis. In Fig. 1e) the invariant cross-section for carbon is shown with no selection in $\cos(\theta_{PROX})$ and in Fig. 1f) it is shown with the condition $\cos(\theta_{PROX})<$0. In Fig. 1f), we observe, consistently with the result of sequential emission of Fig. 1d), a characteristic pattern of a forward sequential emission from the PLF source, well shaped around a Coulomb ring.
No such evidence is shown in Fig. 1e), indicating that the PLF backward sequential decay, going towards the mid-velocity region, is strongly mixed with a non-sequential prompt emission. 

To get more insights to the correlations between isospin, relative velocities and emission time-scale of IMFs we have calculated, for each bin (0.5$\times$0.5 width) in the plane r1-r2 of Fig. 1a), the average N/Z isotopic distributions for all charges between Z=3 and Z=8. Fig. 2 shows, for example, the result for charge Z=6: the largest values of the neutron to proton ratios are obtained for events near the 
diagonal of the V$_{REL}$/V$_{Viola}$(1,3) vs. V$_{REL}$/V$_{Viola}$(2,3) plane,
corresponding to prompt emission (lowest time-scale emission) and to the highest degree of alignment. The neutron content enrichment of the mid-velocity emission with respect to the N/Z of fragments evaporated from PLF or TLF has been previously observed, with different interpretations of the results \cite{mil02,ther05,pia06,pla08}. Our experimental approach introduces two new aspects: i) the $<$N/Z$>$ distribution is not averaged over the whole emission time scale, but it is linked to the reaction dynamics in a consistent way; ii) the $<$N/Z$>$ of the fragments can be correlated with the alignment properties of different emission scenario, so enhancing the experimental sensitivity to select those genuine effects due to isospin dynamics \cite{lio05,bar04}. 

\begin{figure}
\includegraphics[width=0.60\textwidth]{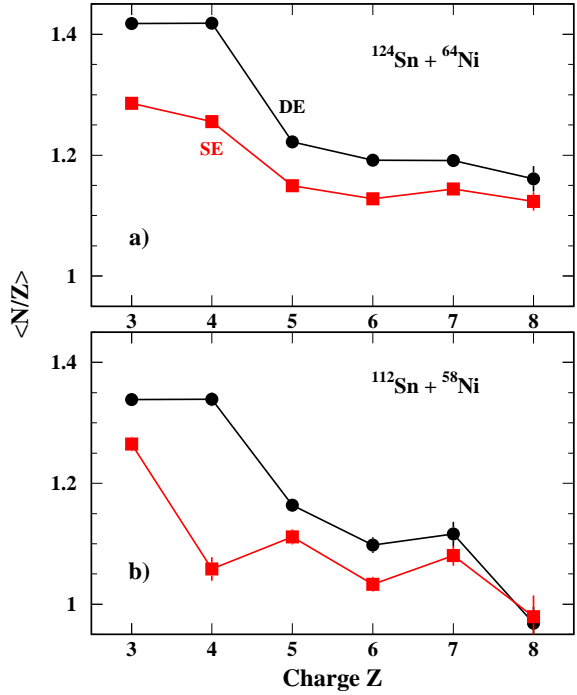}
\caption{(Color online) Experimental $<$N/Z$>$ distribution of IMFs as a function of charge Z for statistical emitted particles (solid squares) and dynamical emitted particles (solid circles), for the reactions a) $^{124}$Sn + $^{64}$Ni and b) $^{112}$Sn + $^{54}$Ni.
}
\label{fig3}
\end{figure}

We show in Figs. 3a) and 3b) the $<$N/Z$>$ as a function of the IMFs charge Z for the reactions 
$^{124}$Sn + $^{64}$Ni and $^{112}$Sn + $^{58}$Ni, respectively. Our purpose is to measure the degree of neutron enrichment at midvelocity and to compare it with the one related with the statistical emission from a PLF source for the two systems with different isospin asymmetry. The condition  $\cos(\theta_{PROX})<$0 has been used to select fragments statistically emitted in the PLF forward hemisphere (solid squares in Fig. 3). Dynamically emitted IMFs (solid circles in Fig. 3) are selected by requiring 
$\cos(\theta_{PROX})>$0.8 and by imposing a further condition in the r$_1$-r$_2$ plane that selects 
events near diagonal of that plane. 
This latter condition is schematically shown by a dashed line in Fig. 2 for charge Z=6. 
We clearly observe that the N/Z ratio for dynamically emitted particles (DE) shows systematically larger values for both systems with respect to the one obtained for statistically emitted particles (SE). A second interesting observation is the flattening of the even-odd effect in the $<$N/Z$>$ distribution of the neutron rich system with respect to the neutron poor one and this effect is present also when comparing the DE distribution respect to the SE one for the neutron poor system. Odd-even effects have been linked to the last evaporation step involving just one neutron or one proton emission \cite{lom11, ric04, ger04} or, more generally, to the last steps in the decay channel 
\cite{cas12}. Both the explanations are closely linked to the ground state binding energy and level density effects of final isotopes \cite{win00,dag11} that are mainly responsible for odd-even effects in evaporation models. Anyway, as it was stated in \cite{dag11} and evidenced mainly in Fig. 3b), a proper selection of the emissing source shows that odd-even effects are influenced by different reaction decay mechanisms. 

\begin{figure}
\includegraphics[width=0.60\textwidth]{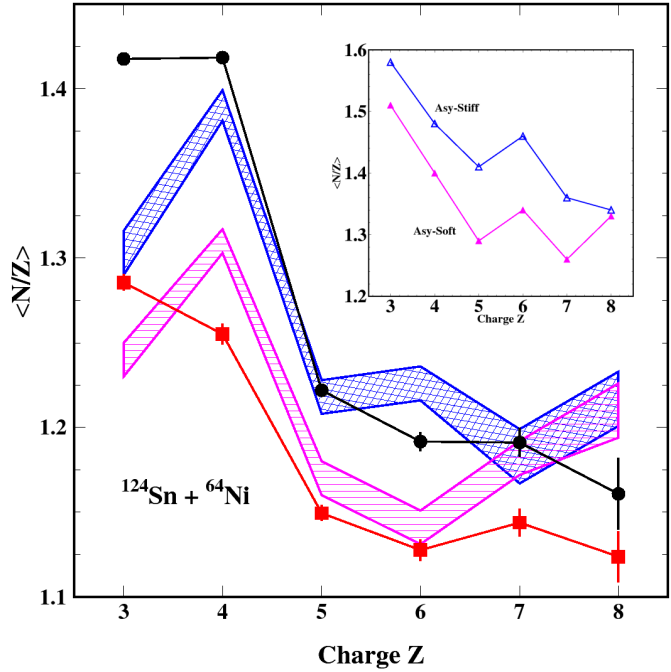}
\caption{(Color online) Experimental $<$N/Z$>$ distribution of IMFs as a function of charge Z for statistically emitted particles (solid squares) and dynamical emitted particles (solid circles), for the reaction $^{124}$Sn + $^{64}$Ni.
Blue hatched area: SMF-GEMINI calculation for dynamical emitted particles and asy-stiff parametrization;
magenta hatched area: asy-soft parametrization. The $<$N/Z$>$ of primary dynamical emitted IMFs 
as a function of their atomic number Z, obtained from SMF calculations, are plotted in the inset.
The hatched zone indicates the error bars in the calculations.}
\label{fig4}
\end{figure}

\section{Comparison with SMF model}
We have compared our data for the neutron rich system to transport theories using the Stochastic Mean Field (SMF) model \cite{dit10,col98}, based on Boltzmann-Nordheim-Vlasov (BNV) equation, already used in \cite{pag04,def05} to describe the basic experimental features of the PLF, TLF, and IMF in ternary reactions. The SMF model implements the nuclear mean field dynamics as well as the effect of fluctuations 
induced by nucleon-nucleon collisions.   
Two different parameterizations of the potential part of the symmetry term of EOS were used. The first one linearly increases with the density (asy-stiff) and
the second one (asy-soft) exhibits a weak variation around the nuclear saturation density $\rho_0$
\cite{dit10}. The corresponding slope parameter $L=3\rho_0(d\epsilon_{sym}/d\rho)_{\rho=\rho_0}$ is around 80 MeV for the asy-stiff and 25 MeV for the asy-soft choices, respectively.
In the current calculation, isovector thermal fluctuations, corresponding to the actual value of the symmetry energy at the neck density, have been implemented \cite{colm}. 
Calculations have been performed at 6 fm impact parameter and by selecting ternary events, as in the experiment. In order to compare the calculations with the data, the primary hot fragments produced by SMF pass throught a de-excitation phase using the statistical model GEMINI \cite{char88}. The average excitation energy of the IMFs before the GEMINI step is E*/A$\approx$ 2.5 A.MeV. 

Fig. \ref{fig4} shows the calculated $<$N/Z$>$ as a function of dynamical emitted IMFs atomic number. As shown in the inset, the two Asy-EoS parameterizations give rather different predictions for the $<$N/Z$>$ of primary fragments: the asy-stiff parametrization  produces more neutron rich fragments respect to the asy-soft choice. After the GEMINI secondary-decay stage the difference in $<$N/Z$>$  between the two parameterizations persists for Z$<$7. The asy-stiff parametrization (blue hatched area) matches the experimental data fairly well.

\begin{figure}
\includegraphics[width=0.60\textwidth]{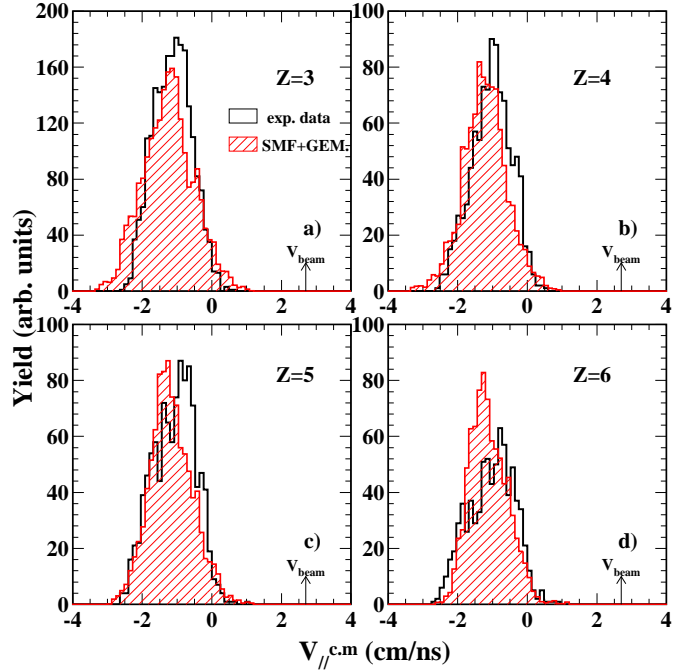}
\caption{(Color online) For the $^{124}$Sn+$^{64}$Ni reaction, (empty, black) experimental velocity spectra 
V$_\parallel$ (in the c.m. system) of dynamically emitted intermediate mass fragments from charge Z=3  
to charge Z=6 and (hashed, red) calculated velocity spectra (SMF+GEMINI) for the same reaction. Calculations have been normalized to data as explained in the text. The arrows indicate the c.m. beam velocity as reference. 
Calculated data are filtered for detector acceptance.}  
\label{fig5}
\end{figure} 

We have checked if the calculation reproduces some basic features of dynamical emitted fragments, like, for example, velocity spectra and charge distributions. 
In Fig. \ref{fig5}a-d) we present the center-of-mass velocity spectra of intermediate mass fragments 
from charge Z=3 to Z=6 selected by requiring the condition $\cos(\theta_{PROX})>$0.8 and by excluding particles that lie along the r$_1$=1 and r$_2$=1 lines. The longitudinal velocity distributions, as 
observed in neck fragmentation, are well centered around the midvelocity region at halfway the velocity 
of the TLF and PLF. In the same figures the experimental data are compared with results of SMF+GEMINI calculations for neck emitted fragments in ternary events
and asy-stiff parametrization (hashed histograms). In the calculated distributions only fragments that are originating from a primary ``neck'' fragment are taken into account. The detector geometry, thesholds and time resolutions have been applied to the calculation, simulating in detail the response and acceptance of the CHIMERA detector (filter). Calculations were 
normalized to data determining the total yield ratio $R=Y(exp)/Y(cal)$ for all charges between Z=3 and Z=9 and assuming the same normalization factor $R$ for each Z. 
Shapes and relative intensities of the experimental longitudinal velocity
distributions are remarkably well reproduced by the SMF+GEMINI simulation: the
calculated distributions, as well the experimental ones, extend up to the
center of mass velocity in the mid-velocity region; it can be also
noticed a slight tendency for the experimental spectra to be peaked at
higher values of V$_\parallel$ respect to the calculated ones at increasing of the
IMF atomic number Z.

\begin{figure}[ht]
\includegraphics[width=0.50\textwidth]{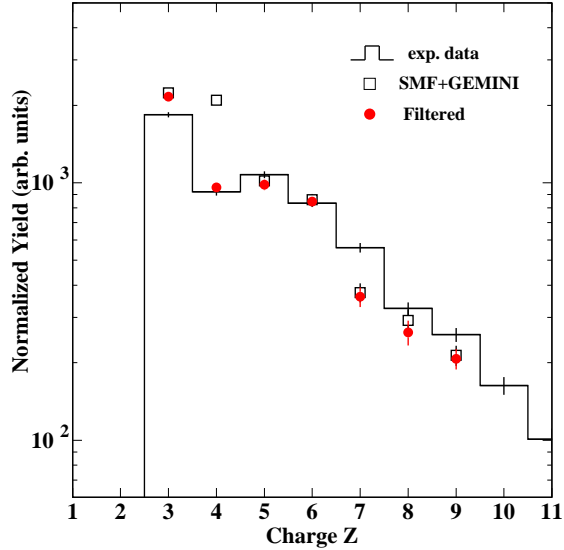}
\caption{(Color online) For the $^{124}$Sn+$^{64}$Ni reaction, experimental charge distribution of dynamically 
emitted IMFs. The squares correspond to the SMF+GEMINI calculation. Red circles: filtered calculated data. The large effect of detector filter on charge Z=4 is due to the unbound $^8$Be isotope that is present in the calculation but filtered because not identified in data analysis.}  
\label{fig6}
\end{figure}

Fig. \ref{fig6} shows the experimental charge distribution for dynamically emitted IMFs compared 
with the calculated charge distribution (square) and filtered by the detector geometry (circles). The same normalization factor $R$ of Fig. \ref{fig5} has been used between experimental and calculated data. 
For charge Z=4, the unbound $^8$Be isotope is not included in the data; thus it has been filtered 
also in the calculations.
The typical exponential behaviour \cite{bar04} of dynamically emitted fragments charge distribution is fairly well reproduced. 

\begin{figure}
\includegraphics[width=0.60\textwidth]{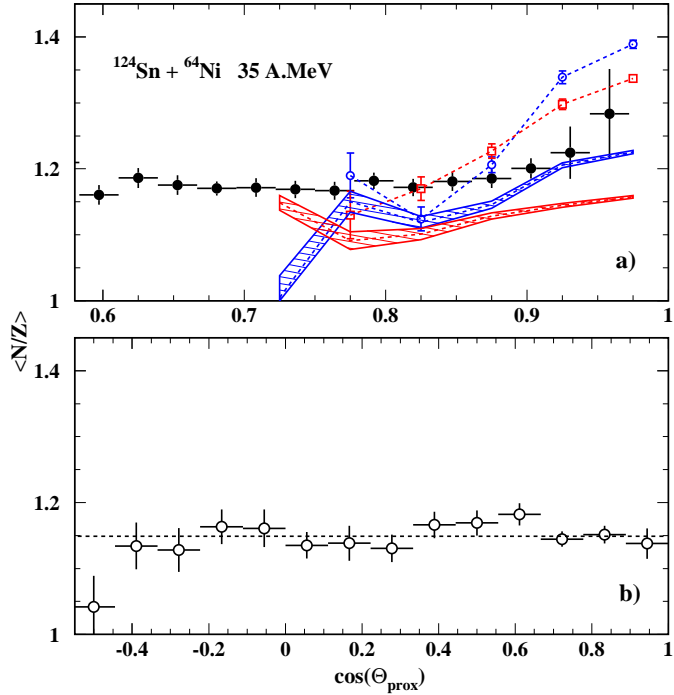}
\caption{(Color online) a) $<$N/Z$>$ as a function of $\cos(\theta_{PROX})$ for charges 5$\le$Z$\le$8: solid circles: experimental points; SMF calculations are shown for primary fragments as empty circles (asy-stiff) and empty squares (asy-soft); SMF-GEMINI calculation for asy-stiff  (blue hatched area) and asy-soft (red hatched area) parametrization. b) experimental correlation for statistically emitted particles (see text). The dashed line gives the average N/Z over the whole $\cos(\theta_{PROX})$ interval.}
\label{fig7}
\end{figure}

Fig. \ref{fig7}a) (solid circles) shows the correlation between $<$N/Z$>$ and $\cos(\theta_{PROX})$ 
for the reaction $^{124}$Sn + $^{64}$Ni. In this case, we have restricted the $<$N/Z$>$ analysis to IMF 
charges between Z=5 and Z=8, excluding lithium and beryllium, in order to increase the sensitivity of the
analysis by avoiding $<$N/Z$>$ strongly reflecting the value of stability line.
We observe an increase of the $<$N/Z$>$ for $\cos(\theta_{PROX})$ values larger than 0.9, corresponding 
to the highest degree of alignment. For comparison Fig. \ref{fig7}b) shows the same correlation obtained 
when only IMFs that lies along r$_1$=1 line in the V$_{REL}$/V$_{Viola}$ plot are taken into account 
(statistical emission). Notice the flat distribution, as expected for a statistical isotropic decay; the average $<$N/Z$>$ value is equal to 1.15.

In Fig. \ref{fig7}a) the SMF calculations of the $<$N/Z$>$ values are plotted as a function of $\cos(\theta_{PROX})$ for the primary dynamically emitted fragments (5$\le$Z$\le$8) with the asy-stiff (open blue circle) and asy-soft (open red squares) parametrizations, respectively. Most of the dynamically emitted fragments are produced at values of $\cos(\theta_{PROX})>$0.8 with increasing value of $<$N/Z$>$ by increasing the degree of alignment. It is interesting to note that the asy-stiff parametrization tends to produce more neutron rich fragments and with a steeper slope with respect to the asy-soft one. Results of SMF+GEMINI are shown in Fig. \ref{fig7}a) as hatched zones for asy-stiff (dark-blue) and asy-soft (grey-red) respectively. Notice that for values of $\cos(\theta_{PROX}$)$>$0.9 the signal of neutron enrichment persists also after the de-excitation stage. Comparison between data and calculation shows, as in 
Fig. \ref{fig4}, that the asy-stiff parametrization matches better the experimental data. These results are consistent with recents measurements obtained from isospin diffusion studies \cite{gal09}, heavy residue production in semi-central collisions \cite{amo09} and transverse collective flow of light charged particles \cite{koh11}. Although final results of $<$N/Z$>$ observable depend also upon the amount of excitation energy given as input in GEMINI calculations, and, consequently, secondary decay tends generally to reduce the sensitivity to symmetry energy \cite{nap10}, we have shown that it is possible to construct robust observable for neck emission dynamics that maintains memory of the early stages of the reaction. This work adds new important observable for isospin dynamics studies in
heavy ion collisions and improves the consistency of the different analyses that have been performed so far on the symmetry term of the EOS \cite{tsa12}. It also opens new perspectives for reaction studies with exotic beams.

\section{Conclusion}
In summary, we have presented new experimental results correlating the emission timescale of intermediate mass fragments (IMF) at mid-rapidity in semi-peripheral collisions with their isotopic composition. We have shown that large values of $<$N/Z$>$ are acquired by light IMFs dynamically emitted in the early stage of the reaction, for both the neutron rich and neutron poor systems studied here. By comparing the experimental data with SMF calculations, we have produced valuable information on the 
parametrization of the symmetry energy term of EOS, getting indication for a moderately stiff 
symmetry potential, and new constraints for further simulations of the reaction dynamics. 

\section*{Acknowledgments}
We thank the INFN-LNS accelerator staff for providing beam of excellent quality. We 
acknowledge also C. Marchetta and E. Costa for providing high-quality targets. 
This work for V. Baran was partially supported by the Romanian National Authority 
for Scientific Research, CNCS-UEFISCDI, Project No. PN-II-ID-PCE-2011-3-0972.

\end{document}